\documentclass[a4paper, 11pt]{article}
\usepackage{pos}

\usepackage{graphicx}
\usepackage{amsmath}
\usepackage{amsfonts}
\usepackage{hyperref}
\usepackage{color}
\usepackage{placeins}
\usepackage{tabu}
\usepackage{nicefrac}
\usepackage{ulem}
\usepackage{float}
\usepackage{subfig}
\usepackage{multirow}
\usepackage{tikz-feynman}
\newcommand{\orcid}[1]{\,\href{https://orcid.org/#1}{\includegraphics[width=9pt]{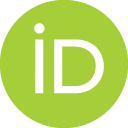}}\,}
 
\newcommand{\chidof}{\chi^2/N_{dof}}

\newcommand{\orcidPD}{0000-0001-7960-7953} %
\newcommand{\orcidHP}{0000-0001-8815-4255} %
\newcommand{\orcidVG}{0000-0002-2393-8507} %
\newcommand{\orcidIH}{0000-0003-1998-038X} %

\FullConference{31st International Workshop on Deep Inelastic Scattering (DIS2024)\\
 8–12 April 2024\\
Grenoble, France\\}

\abstract{
We present numerical studies of the leading non-linear corrections to the DGLAP evolution equations of parton distribution functions (PDFs) resulting from gluon recombination, which reduce the pace of evolution at small momentum fractions $x$. The non-linear evolution is implemented in the \textsc{HOPPET} and \textsc{xFitter} toolkits and used to carry out fits of proton PDFs using lepton-proton deep inelastic scattering data from HERA, BCDMS and NMC. While we do not find evidence for non-linear effects, we are able to set upper limits for their strength. We also quantify the potential impact of longitudinal structure function measurements at the Electron-Ion Collider and the Large Hadron Electron Collider on future fits.
}

\begin{document}
\title{Global fits of proton PDFs with non-linear corrections from gluon recombination}

\author{P.~Duwent\"aster\orcid{\orcidPD}}
\emailAdd{pitduwen@jyu.fi}
\affiliation{University of Jyvaskyla, Department of Physics, P.O. Box 35, FI-40014 University of Jyvaskyla, Finland}
\affiliation{Helsinki Institute of Physics, P.O. Box 64, FI-00014 University of Helsinki, Finland}

\author{V. Guzey\orcid{\orcidVG}}

\author{I. Helenius\orcid{\orcidIH}}

\author{H. Paukkunen\orcid{\orcidHP}}

\date{\today}

\maketitle
\section{Introduction}
\label{sec:intro}

In the framework of collinear factorization~\cite{Collins:1989gx}, the cross sections for hard processes involving intial-state hadrons are calculated as convolutions of non-perturbative, process-independent parton distribution functions (PDFs) $f_i(x,Q^2)$ and perturbative, process-dependent coefficient functions. 
The PDFs describe the density of partons (gluons and quarks) of flavor $i$ carrying a fraction $x$ of the hadron's momentum at a resolution scale $Q^2$. The $Q^2$ dependence is given by the Dokshitzer-Gribov-Lipatov-Altarelli-Parisi (DGLAP) evolution equations \cite{Dokshitzer:1977sg,Gribov:1972ri,Gribov:1972rt,Altarelli:1977zs}, which results in an increasing density of partons carrying a smaller momentum fraction as the scale $Q^2$ increases. However, as the parton densities become sufficently high, the recombination of partons becomes important and limits the growth of PDFs at small $x$ as $Q^2$ increases. 

The Gribov-Levin-Ryskin-Mueller-Qiu (GLR-MQ) corrections~\cite{Qiu:1986wh, Kovchegov:2012mbw, Mueller:1985wy, Gribov:1983ivg} to the DGLAP evolution equations describe this recombination in a leading logarithmic approximation in $Q^2$ and $1/x$. However, due to the leading logarithmic approximation in $1/x$, the GLR-MQ equations violate the momentum conservation. A derivation of recombination corrections without approximation in $1/x$ has been presented by Zhu and Ruan in Refs.~\cite{Zhu:1998hg,Zhu:1999ht}. For dimensional reasons, these higher-twist effects imply a presence of a dimensionful parameter $R$, which controls the strength of the non-linear effects. Correspondingly, the inverse of this parameter $1/R = Q_r$ gives the momentum scale below which the effects of recombination will be considerable. 

To study the consequences of the resulting evolution, we have extended the \textsc{HOPPET} evolution code~\cite{Salam:2008qg} with these non-linear corrections to perform numerical studies and coupled it to the \textsc{xFitter} framework~\cite{xFitter:2022zjb, Alekhin:2014irh} to perform global fits of proton PDFs.

\begin{figure*}
	\centering
	\includegraphics[width=0.7\textwidth]{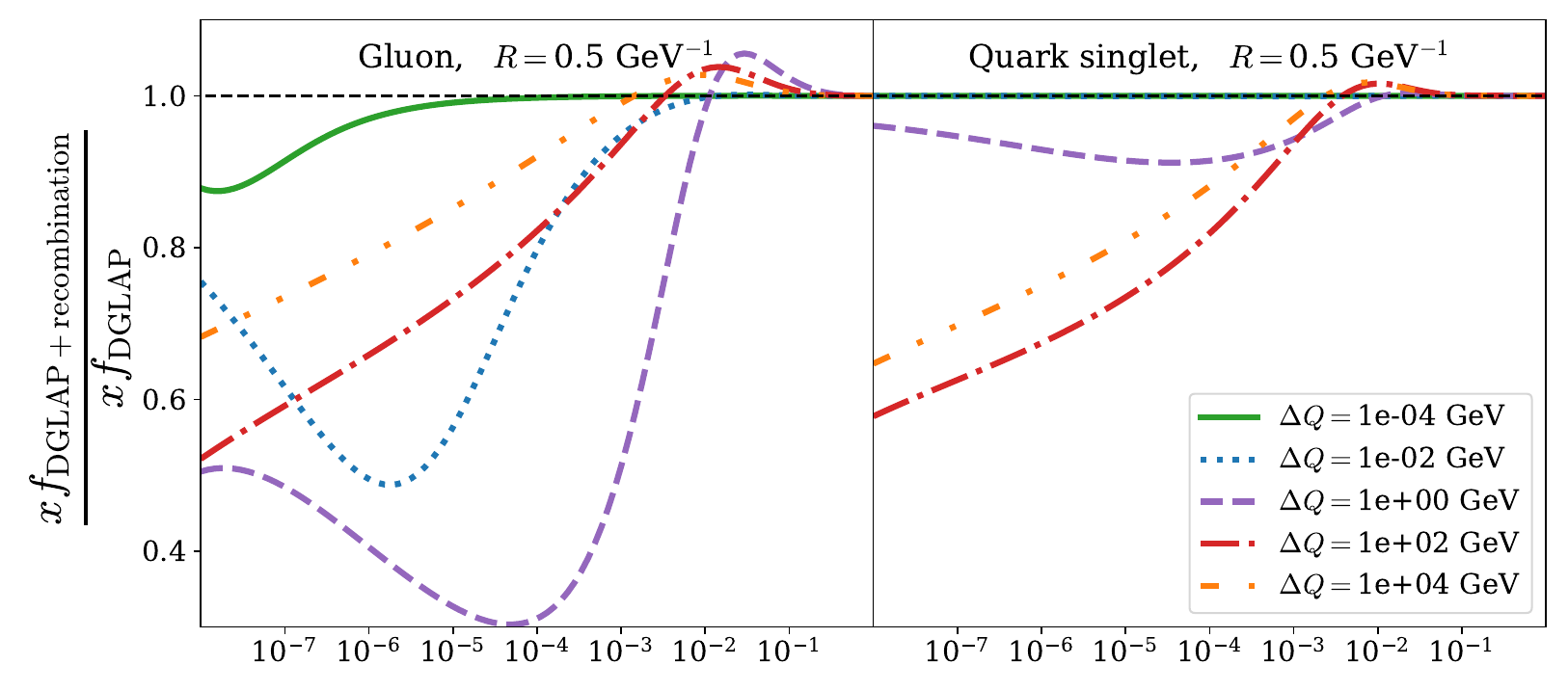}
	\caption{Ratios between the PDFs evolved using the non-linear and linear evolution equations with the CJ15 initial condition at $Q_0=1.3\,{\rm GeV}$ as a function of $x$ and at different values of $Q=Q_0+\Delta Q $. The left column corresponds to the gluon PDF, while the right column shows the quark singlet  distribution.}
    \label{fig:qdep}
\end{figure*}

In Fig.~\ref{fig:qdep} we compare regular next-to-next-to-leading order (NNLO) DGLAP evolution to evolution with non-linear corrections by showing the ratios between PDFs evolved using the non-linear and linear evolution as a function of $x$ at different values of $Q$. At the initial scale of $Q_0 = 1.3$\,GeV both PDFs are given by the CJ15~\cite{Accardi:2016qay} proton fit. 
The parameter $R$ is set to 0.5\,GeV$^{-1}$ and the $Q$ values are chosen as $Q_0 + \Delta Q$ with $\Delta Q \in \{10^{-4}, 10^{-2}, 1, 10^2, 10^4\}$\,GeV. One can clearly see a dynamically generated suppression at small $x$ and a corresponding enhancement at intermediate $x$ for both the gluon and the quark singlet PDFs. 
While the relative effects of non-linearities first increase as $Q$ increases, there is eventually a turnaround and the differences begin to decrease. This is due to the fact that $G^2(x,Q^2)/Q^2 \rightarrow 0$ as $Q^2 \rightarrow \infty$ so that the non-linear terms die away at asymptotically large $Q^2$. 

\section{PDF fits with recombination effects}
\label{sec:fit}
To obtain a more complete and consistent analysis of the impact of non-linear corrections, we perform a new global PDF fit with the corrections included throughout the fitting process. To perform these fits, we use the \textsc{xFitter} package~\cite{xFitter:2022zjb, Alekhin:2014irh}, which we couple to our modified version of \textsc{HOPPET}. The fits are performed at NNLO accuracy in both evolution and scattering coefficients. To make sure that our findings are independent of the assumed parameterization for the PDFs, we repeat the full analysis with two different gluon parameterizations, which we refer to as parameterization 1 and 2. Parameterization 1 exactly mirrors the parameterization used in the HERAPDF2.0 analysis~\cite{H1:2015ubc}, where the gluon distribution tends to become negative at small values of $x$ and $Q^2$. In the case of parameterization 2 we apply the same parameterization for gluons as we have for quarks, which enforces positivity for the gluon at the initial scale. The parameter $R$ is kept fixed in each fit and the procedure is repeated for $R$ values from 0.2\,GeV$^{-1}$ to 3.0\,GeV$^{-1}$. For $R\gtrsim 3.0$\,GeV$^{-1}$, the non-linear modifications were found to be negligible. We determine the uncertainties of the obtained PDFs by using the Hessian method~\cite{Pumplin:2001ct} with a tolerance of $\Delta \chi^2_{\mathrm{max}} = 20$ at the 90\% confidence level.

Our analysis includes lepton-proton DIS data from the BCDMS, HERA and NMC experiments with a kinematic cut of $Q^2>3.5$\,GeV$^2$. To investigate whether the inclusion of non-linear corrections can help to alleviate the tensions between usual NNLO fits and data in the low-$Q^2$ region, we also perform a set of fits with parameterization 1 and a cut of $Q^2>1$\,GeV$^2$. %
For full details of the fitting setup and results, see Ref.~\cite{Duwentaster:2023mbk}.

\begin{figure}
    \centering
    \includegraphics[width=0.5\textwidth]{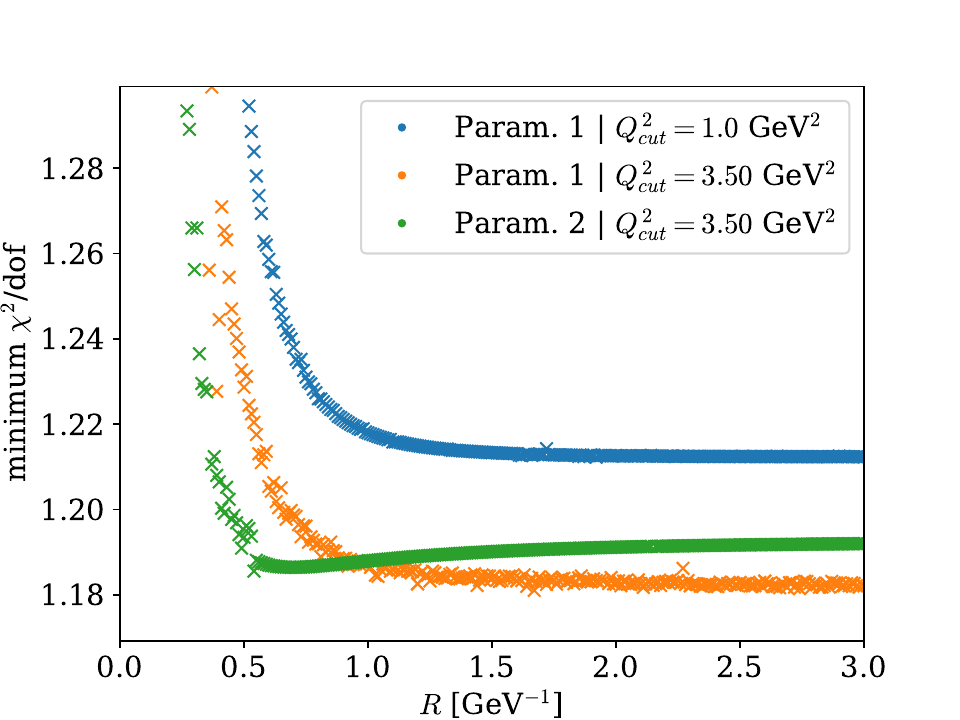}
    \caption{Minimum $\chidof$ as a function of $R$ for parameterization 1 with $Q^2 > 3.5$\,GeV$^2$  (orange) and $Q^2 > 1$\,GeV$^2$ (blue), and parameterization 2 with $Q^2 > 3.5$\,GeV$^2$ (green). 
    }
    \label{fig:Chi2Minima}
\end{figure}

\begin{figure}
    \centering
    \includegraphics[width=0.8\textwidth]{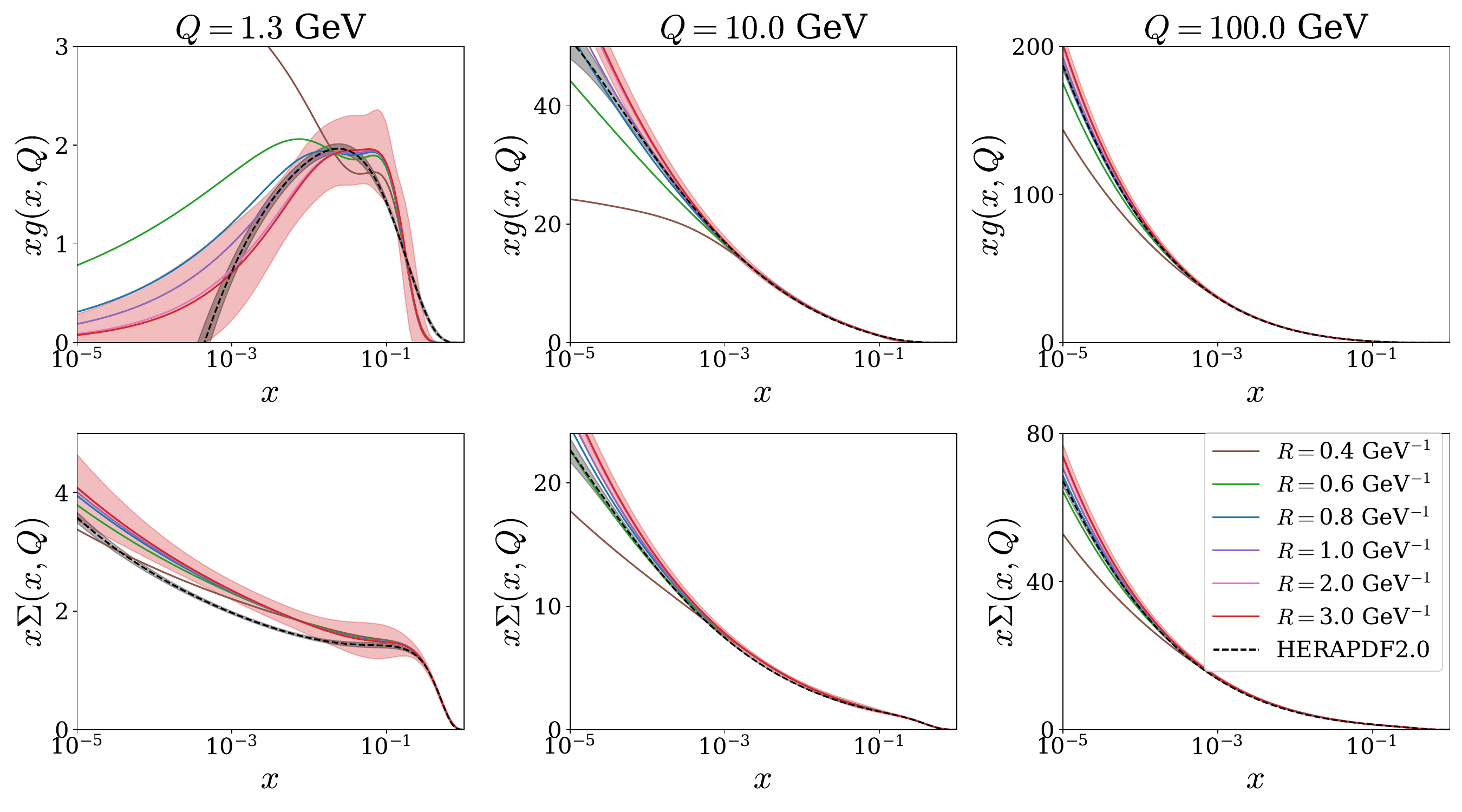} 
    \caption{PDFs resulting from fits at various $R$ values. The top row shows the gluon PDF, while the bottom one shows the quark singlet PDF. The columns correspond to different scales $Q$.}
    \label{fig:PDFcomp_b}
\end{figure}

Fig.~\ref{fig:Chi2Minima} shows the resulting $\chidof$ values for the minimum of each fit against the different $R$ values. 
When using parameterization 1, the resulting $\chidof$ values generally decrease with increasing $R$, i.e., with weaker non-linear corrections, but change only by an insignificant amount for $R>1$\,GeV$^{-1}$. Down towards values of  $R\approx0.7$\,GeV$^{-1}$, the $\chidof$ increases slightly, but below that it diverges rapidly. Lowering the cut from $Q^2>3.5$\,GeV$^2$ to $Q^2>1$\,GeV$^2$ does not change the shape of the curve, but moves it upwards with an average $\chi^2$ of 2 for each added data point, indicating that the non-linear corrections do not help to alleviate the tensions with lower-$Q^2$ data. 
When using parameterization 2, the best description of the data is obtained with $R$ values in the interval $0.5 < R < 0.7$\,GeV$^{-1}$ although the difference with respect to $R\rightarrow\infty$ is small and the minimum is still slightly higher than the best fit achieved with parameterization 1. This implies that the local minimum in $\chidof$ is not due to the physical effect of recombination, but rather just the added flexibility from an additional parameter to an otherwise insufficiently flexible parameterization. 
The proton PDFs obtained in the fits with the $Q^2>3.5$\,GeV$^2$ cut and parameterization 2 for a representative selection of $R$ values are shown in Fig.~\ref{fig:PDFcomp_b}. The top panels show the gluon distribution at $Q=\{1.3 , 10, 100\}\,{\rm GeV}$ and the bottom panels show the quark singlet distribution. The fits are compared to HERAPDF2.0 with uncertainties at the 90\% confidence level shown only for HERAPDF2.0 and $R=3.0$\,GeV$^{-1}$. %
At intermediate and high $x$, our PDF fit with $R=3$\,GeV$^{-1}$ agrees well with HERAPDF2.0, but has larger uncertainties due to the higher Hessian tolerance. At lower $x$, differences due to the parameterization can be observed at the initial scale, where the gluon of HERAPDF2.0 dips below 0 for $x<5\cdot10^{-4}$. Comparing our fits at the initial scale, the highest $R$ value leads to the lowest gluon PDF at small $x$, while lower values of $R$ lead to increasingly large gluon PDFs. After short evolution in $Q^2$, however, this ordering quickly inverts due to the recombination effects at lower $R$ values leading to stronger suppression of the small-$x$ gluon. Towards even higher $Q$ values, the PDFs move closer together again, as the recombination terms are suppressed by $1/Q^2$. The systematics of the $R$ dependence are the same in the fits with parameterization 1, which gives us confidence that the findings are not sensitive to the choice of parameterization. %

\begin{figure}
    \centering
    \includegraphics[width=0.98\textwidth]{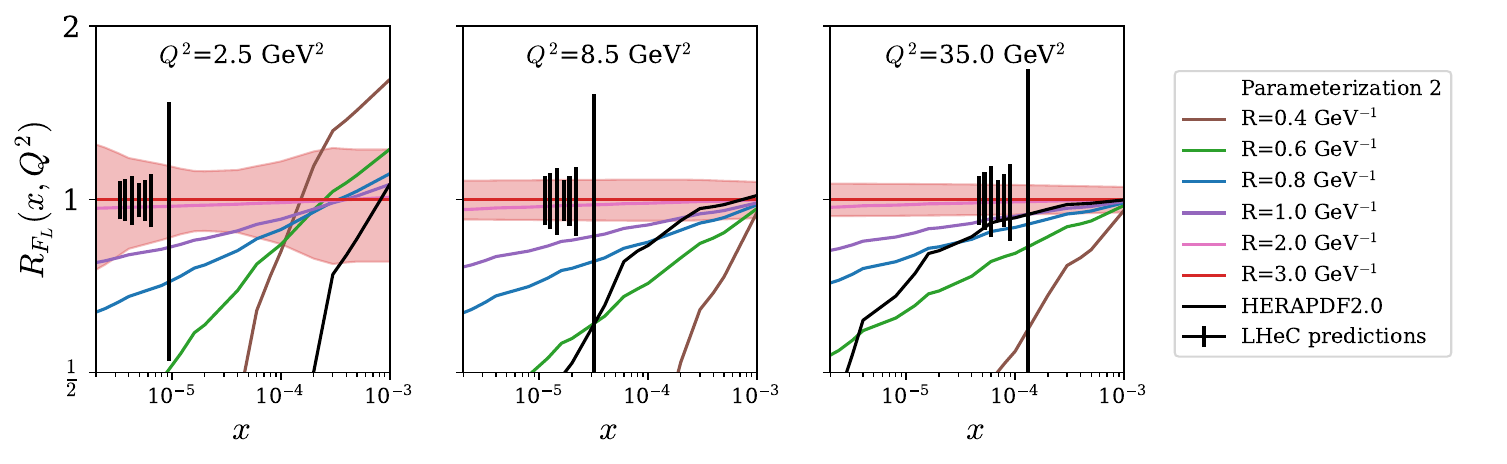}
    \caption{The ratios of the longitudinal structure functions corresponding to different values of $R$ to that with $R=3$\,GeV$^{-1}$ as a function of $x$.}
    \label{fig:LHeC_FL_2}
\end{figure}

The longitudinal structure function $F_L$ is directly sensitive to the gluon PDF and is therefore expected to be very sensitive to non-linear effects. Therefore, future DIS experiments such as the EIC~\cite{Accardi:2012qut} and LHeC~\cite{LHeC:2020van, LHeCStudyGroup:2012zhm} could put tighter constraints on the presence of these effects. To investigate this, we compare the spread of $F_L$ predictions by taking the ratio of the predictions resulting from each of the PDFs fitted with different $R$ values over that at $R=3$\,GeV$^{-1}$. These ratios can then be compared to projected relative uncertainties of future $F_L$ data from the LHeC White Paper~\cite{LHeC:2020van} in Fig.~\ref{fig:LHeC_FL_2}. The projected experimental uncertainties are shown as black error bars. This gives us an estimate of whether future experiments can constrain the strength of non-linear effects under the assumption that $F_L$ measurements can be accurately reproduced by the theory. %
The reach towards lower $x$ values at the LHeC would put significantly tighter constraints on non-linear effects, since the different $R$ values lead to differences in predictions that are much larger than the projected uncertainties in this kinematic region.
The same analysis was performed for EIC predictions in Ref.~\cite{Duwentaster:2023mbk}, but the projected EIC data does not reach as far into the $x$ region, where the different $R$ values produce different results. 

\section{Conclusions and Outlook}
\label{sec:conclusions}
We have presented numerical studies of DGLAP evolution with non-linear gluon recombination corrections as derived by Zhu and Ruan. These corrections improve upon the commonly used GLR-MQ equations by omitting the leading logarithmic approximation in $1/x$ and thereby restoring the PDF momentum sum rule. We have implemented these corrections in the \textsc{HOPPET} and \textsc{xFitter} toolkits to perform global fits of proton PDFs to BCDMS, HERA and NMC data on DIS with varying values of the dimensionful parameter $R$. We found that both parameterizations result in a good description of the data for large values of $R$, i.e. small or no non-linear effects. The fit quality eventually deteriorates in the case of strong non-linear effects with $R < 0.4$\,GeV$^{-1}$ being excluded by the data, which translates to an upper limit for the recombination scale $Q_r = 1/R \lesssim 2.5 \, {\rm GeV}$. We show that future data taken at the EIC and particularly at the lower $x$ values reachable at the LHeC could provide stronger constraints on the strength of non-linear effects.

The tools developed in this work and LHAPDF files~\cite{Buckley:2014ana} for the fitted PDFs are available here: \url{https://research.hip.fi/qcdtheory/nuclear-pdfs/}.

\section*{Acknowledgements}
Our work has been supported by the Academy of Finland, projects 331545 and 330448, and was funded as a part of the Center of Excellence in Quark Matter of the Academy of Finland, project 346326. This research is part of the European Research Council project ERC-2018-ADG-835105 YoctoLHC. We acknowledge grants of computer capacity from the Finnish Grid and Cloud Infrastructure (persistent identifier \mbox{urn:nbn:fi:research-infras-2016072533}).

\bibliographystyle{utphys}
\bibliography{refs.bib,extra.bib}

\end{document}